\documentclass[runningheads]{svmult}

\usepackage{makeidx}   % allows index generation
\usepackage{graphicx}  % standard LaTeX graphics tool
\usepackage{subeqnar}  % subnumbers individual equations
\usepackage{multicol}  % used for the two-column index
\usepackage{physprbb}  % modified textarea for proceedings,
\makeindex             % used for the subject index
\begin{document}
\title*{Observable Effects of Shocks in Compact and Extended Presupernovae}
\toctitle{Observable Effects of Shocks
\protect\newline in Compact and Extended Presupernovae}
% allows explicit linebreak for the table of content
%
%
\titlerunning{Effects of Shocks in Presupernovae}
% allows abbreviation of title, if the full title is too long
% to fit in the running head
%
\author{Sergey Blinnikov\inst{1}
\and Nikolai Chugai\inst{2}
\and Peter Lundqvist\inst{3}
\and Dmitrij Nadyozhin\inst{1}
\and Stan Woosley\inst{4}
\and Elena Sorokina\inst{5}}
\authorrunning{Sergey Blinnikov et al.}
% if there are more than two authors,
% please abbreviate author list for running head
%
%
\institute{ITEP, 117218, Moscow, Russia, and MPA, Garching, Germany
\and Institute of Astronomy, RAS, 109017 Moscow,  Russia
\and %Stockholm Center for Physics, Astronomy and Biotechnology,
Stockholm Observatory, AlbaNova, SE-106~91 Stockholm, Sweden
\and UCO/Lick Observatory,  University of California, Santa Cruz, CA 95064, USA
\and Sternberg Astronomical Institute, 119992 Moscow, Russia, and MPA, Garching }

\maketitle              % typesets the title of the contribution

\begin{abstract}
 We simulate shock propagation in a wide range of core-collapsing
presupernovae: from compact WR stars exploding as SNe~Ib/c through very
extended envelopes of the narrow-line SNe~IIn.
We find that the same physical phenomenon of radiating shocks
can produce outbursts of  X-ray radiation
(with photon energy $3kT \sim 1$ keV) lasting only a second in SNe~Ib/c,
as well as a very high flux of visual light, lasting for months, in SNe~IIn.
\end{abstract}

\section{Introduction}

Shock waves created at supernova explosions are observed mostly
at the stage of supernova remnants due to their x-ray emission.
For supernovae themselves shocks are not observed directly:
normally the stage of shock break-out through the presupernova
surface layers produces a short-lived transient of hard radiation.
Yet those transients may have important observational consequences.
Moreover, in some supernovae
the shock propagation in surrounding CSM may be
decisive in producing their light on time-scales from days to years. %of months.

A shock propagating
down the profile of decreasing density should accelerate
\cite{blinn.GFK56Sak60}.
The acceleration ends when the leakage %diffusion
of hard photons from the shock
front into outer space becomes efficient enough.
We simulate shock propagation in a variety of core-collapsing presupernova
models using a code with hydrodynamics coupled to multi-energy-group time-dependent
radiation transfer. In our previous work on extended type II SNe
\cite{blinn.Blin93j}, with radii of a few hundred $R_\odot$, we have found the
peak effective temperature at the shock break-out, $T_\mathrm{eff}\sim
1.5\times 10^{5}$K. For SN 1987A, with its presupernova radius
of only  $\sim 50 R_\odot$, we have got  $T_\mathrm{eff}\sim 6\times 10^{5}$K
\cite{blinn.Blinn87a}.
Due to effects of scattering the maximum color temperature is a factor of
2-3 higher.

%It was realized long ago that

In compact presupernovae, like SNe~Ib/c,
the shock can become relativistic \cite{blinn.JM71} and
is able to produce a burst of X-ray and even $\gamma$-ray
radiation  \cite{blinn.Col69,blinn.BKINC75,blinn.WooAA93}.

%The case
%of a peculiar type Ic SN1998bw, probably related to GRB980425,
%has shown that SNe~Ib/c may have the highest explosion energy
%and highest production of $^{56}$Ni among core-collapsing supernovae
%\cite{blinn.Iwamoto,blinn.WooES99}. % ,blinn.Nakamura}.
Supernovae of type Ib/c are also interesting for theory due to
problems with their light curve and spectral modeling. Their understanding
may serve as diagnostics of mass loss from massive stars at
the latest phases of evolution.

\section{Compact presupernovae: SNe~Ib/c}
%
% In compact presupernovae (on order of one solar radius or
%less), like SNe~Ib/c, the shock can become relativistic and is
%able to produce a short burst of X-ray radiation. We present
%predictions for X-ray light for these events.
%
%Supernovae of type Ib/c are interesting due to
%\begin{itemize}
%\item
%probably the highest explosion energy
%\item
%highest production of $56$Ni in core-collapsing SNe
%\item
%probable relation to (some) GRBs
%\item
%problems with light curve modeling
%\item
%diagnostics of mass loss for massive stars
%\end{itemize}
%
%Improvement in the theory of shock breakout in SNe Ib/c in
%the current work:
%\begin{itemize}
%\item
%multi-energy-group time-dependent radiation transfer
%\item
%taking into account of (some) relativistic effects
%\item
%predictions of X-ray flash spectra for future missions
%\end{itemize}
%

Numerical modeling of shock breakout in SNe Ib/c
was done previously using some simplifying
approximations \cite{blinn.ensmanphd}. Our  method, realized
in  the code  {\sc stella}, allows us to get more reliable
predictions for the outburst.
Improvements in the theory done in
the current work are: multi-energy-group time-dependent radiation transfer, and
taking into account of (some) relativistic effects.

A representative presupernova in our runs was a WR star built by the
code {\sc kepler} \cite{blinn.woolangw2} (model 7A).
Late light curves and spectra for this
model were studied in  \cite{blinn.Wee97}.
%We present UBV light curves found for the same model
%by {\sc stella} in Fig.~\ref{blinn.mags62l}.
%\begin{figure}[ht]
%   \centerline{\includegraphics[width=.6\textwidth]{mags62l.ps}}
%  \caption{Theoretical UBV fluxes for the model 7A, exploded
%  with $E_{\rm kin}=1.47\times 10^{51}$ erg in comparison
%    with observations of SN1962L of type Ib.}
%  \label{blinn.mags62l}
%\end{figure}

The Model 7A has mass 3.199 $M_\odot$ (including the mass of the
collapsed core). Its radius prior to explosion is not strictly
fixed because the outer mesh zones in {\sc kepler} output
actually model the strong stellar wind and are not
in hydrostatic equilibrium.
So,  we fixed the radius by hand and
we got four models with radii from 0.76 up to 2 $R_\odot$\ which  {\em are}
in hydrostatic equilibrium. Our results are summarized in the Table.

\begin{figure}[ht]
  \centerline{\includegraphics[width=0.45\textwidth]{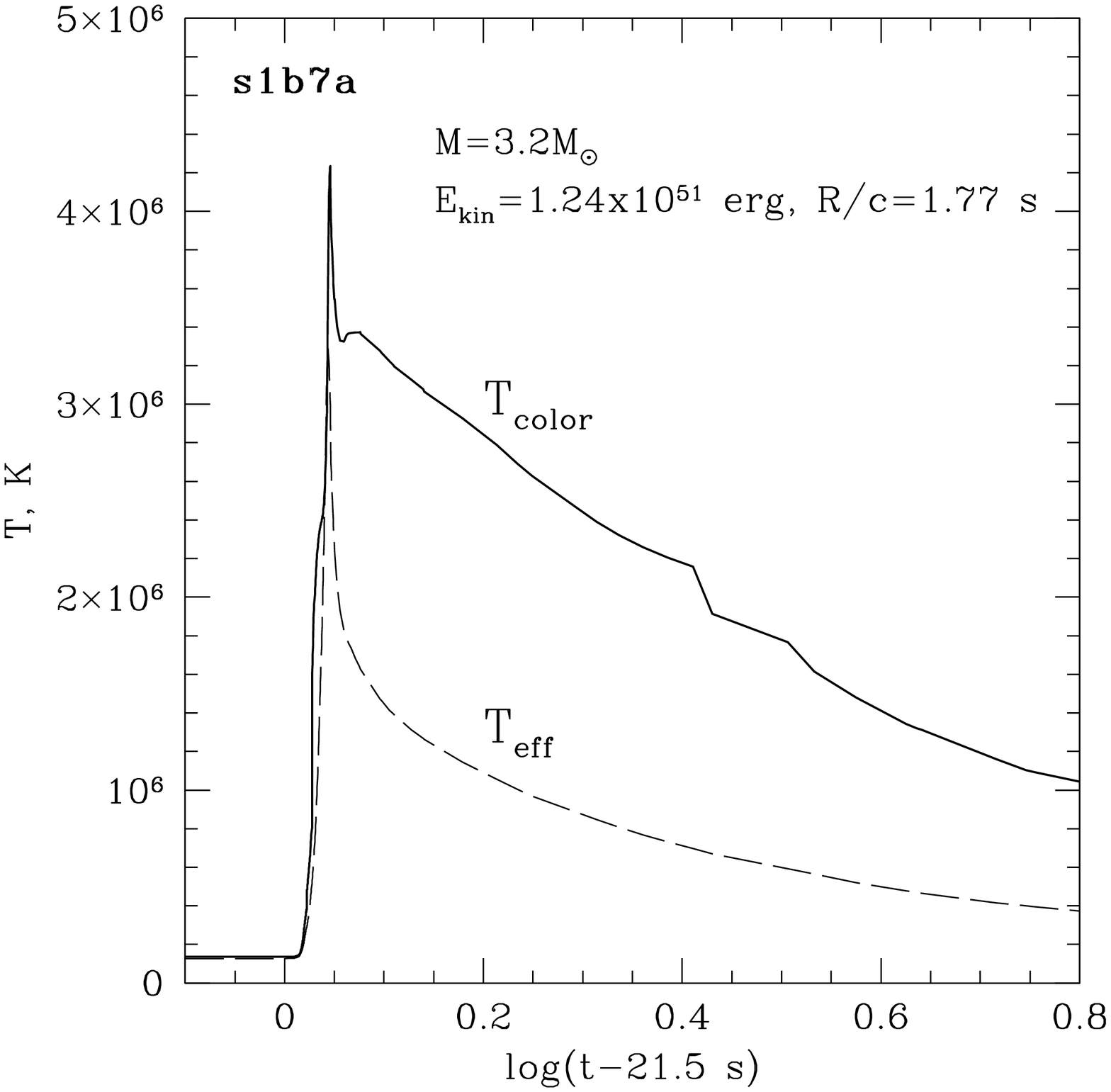} %{lTbTe.ps}
         \includegraphics[width=0.45\textwidth]{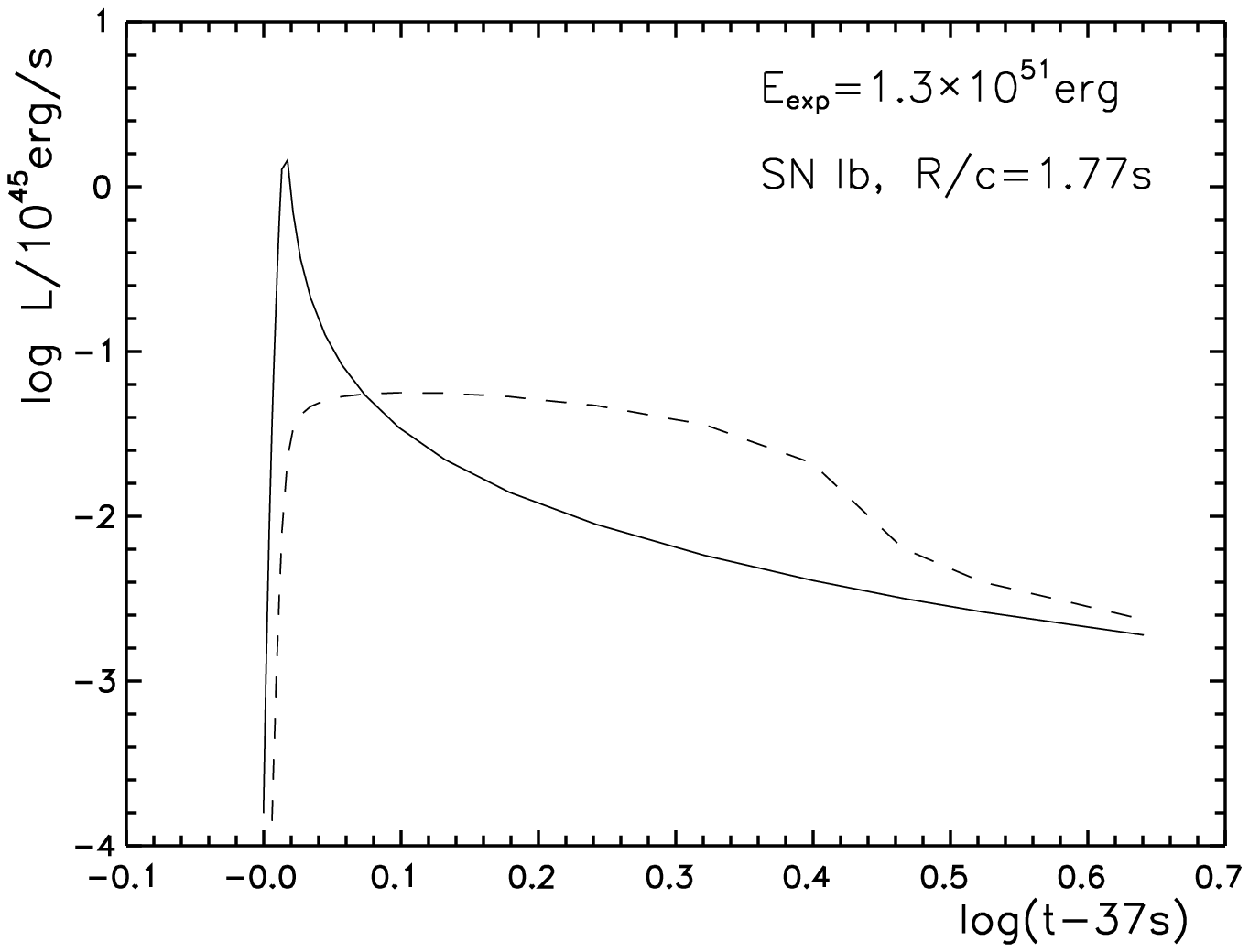}} % swblibqz.eps}}
  \caption{Effective and color temperatures of emerging
  radiation (left).
 Shock breakout
luminosity  found by the hydrocode {\sc snv} (right)
Dashed lines demonstrate the effect
of averaging the light curve due the light travel time correction}
  \label{blinn.lTbTe}
\end{figure}

Explosions with {\sc kepler} \cite{blinn.Wee97}
gave a maximum temperature of photons $T \sim 5\times 10^5$ K
at shock breakout. We have much finer mesh zoning at the edge of
the star (down to $1\times 10^{-12} M_\odot$) and better physics and we find
much higher values of $T$. The left plot in Fig.~\ref{blinn.lTbTe} shows
the difference between effective and color temperatures (labels `e'
and `c', respectively, in the Table).
One should note that the peak values of
luminosity and temperature given in the table and on the left plot
of Fig.~\ref{blinn.lTbTe} do not contain the
light travel time correction. Its effect on the luminosity, i.e.
smearing the peak on the time-scale of $\sim R/c$ is shown on the
right plot of Fig.~\ref{blinn.lTbTe}.
%\begin{figure}[ht]
% \centerline{\includegraphics[width=0.45\textwidth]{swblibqz.eps}
%  \includegraphics[width=0.45\textwidth]{llumavg.ps}}
%  \caption{The left figure shows the results for shock breakout
%  luminosity  by the equilibrium radiation diffusion hydrocode {\sc snv};
%  the right one: a similar presupernova model, results obtained by
%  {\sc stella}. Dashed lines show the light travel time corrected
%  luminosity.}
%  \label{blinn.llumavg}
%\end{figure}
Models, labeled as 3.5N were computed by one of us (D.K.N.) in 1992
with the equilibrium radiation diffusion hydrocode {\sc snv}.
%Left plot in Fig.~\ref{blinn.llumavg} shows one of his results.
%Dashed lines in Fig.~\ref{blinn.llumavg} demonstrate the effect
%of averaging the light curve due the light travel time correction.
%
\begin{table}
\caption{Parameters of shock breakouts}
\begin{center}
\renewcommand{\arraystretch}{1.4}
\setlength\tabcolsep{15pt}
%\begin{tabular}{@{}llp{1.8cm}lll}
\begin{tabular}{llllll}
\hline\noalign{\smallskip}
 $M^{\mathrm a}$ & $R_0^{\mathrm a}$ & $E_{\rm kin}^{\mathrm b}$ &
$L_{\rm p}^{\mathrm c}$   & $T_{\rm p}^{\mathrm c}$ & $\Delta t^{\mathrm d}$  \\
 $M_\odot$ & $R_\odot$ & foe & erg/s   & $10^{6}$K & s  \\
\noalign{\smallskip}
\hline
\noalign{\smallskip}
3.2  &  0.76 &  1.24   & $4.2\times 10^{44}$  & 4.2c  &  0.021 \\ % s1b7a763
3.2  &  1.00 &  1.32   & $5.8\times 10^{44}$  & 4.3c  &  0.026   \\ % s1b7a1w15
3.2  &  1.23 &  1.30   & $6.8\times 10^{44}$  & 4.3c  &  0.043   \\ % s1b7a1t2
3.2  &  2.   &  1.39   & $9.4\times 10^{44}$  & 4.3c  &  0.12   \\ % s1b7wl15
3.2  &  2.   &  4.36   & $3.6\times 10^{45}$  & 5.3c  &  0.028   \\ % s1b7wl18
3.2  &  2.   &  8.86   & $4.8\times 10^{45}$  & 7.2c  &  0.020   \\ % s1b7wl16
\hline
3.5N  &  0.76 &  1.30   & $1.4\times 10^{45}$  & 5.1e  &  0.028 \\ % Nad R0.763
3.5N  &  1.23 &  1.30   & $8.1\times 10^{44}$  & 3.5e  &  0.067  \\ % Nad R1.23
\noalign{\smallskip}
\hline
\noalign{\smallskip}
\end{tabular}
\end{center}
$^{\mathrm a}$
presupernova mass and radius, respectively,
           in solar units. \\
$^{\mathrm b}$ kinetic energy at infinity in $10^{51}$
ergs.\\
$^{\mathrm c}$ peak luminosity and temperature.\\
$^{\mathrm d}$ the width of the
            light curves at 1 stellar magnitude below $L_p$.
\label{Tab1}
\end{table}

%\newpage

\section{Shocks in CSM in Type IIn supernovae}

The narrow-line Type II supernovae (SNe IIn) are embedded in massive
circumstellar shells (wind) extending from tens of thousands solar radii
(SN~1998S) to $\sim 10^{17}$ cm (SN~1988Z). %reaching tens of thousands solar radii.
One of
the brightest SNIIn, SNII 1994W in NGC 4041, displays a spectrum dominated for $\sim 200$ days
by low-ionization P-Cygni lines with widths of order 10$^3$ km/s,
and at late times by narrower H$_\alpha$ in pure emission.
The lightcurve shows a plateau for $\sim 120$ days, after
which the luminosity drops by $\sim 3.5$ magnitudes in $V$ in only 12 days.
The existence of such a drop, caused by circumstellar shells,
was emphasized in paper \cite{blinn.GrNad87}.
%The observations of SN 1994W are tied  into a coherent model
The interpretation of spectra and light curve of SN 1994W
leads to a coherent model in which the supernova interacts
with a massive circumstellar shell ejected roughly 2 years prior
to the explosion \cite{blinn.Chugai}.

\begin{figure}[ht]
  \centerline{\includegraphics[width=0.45\textwidth]{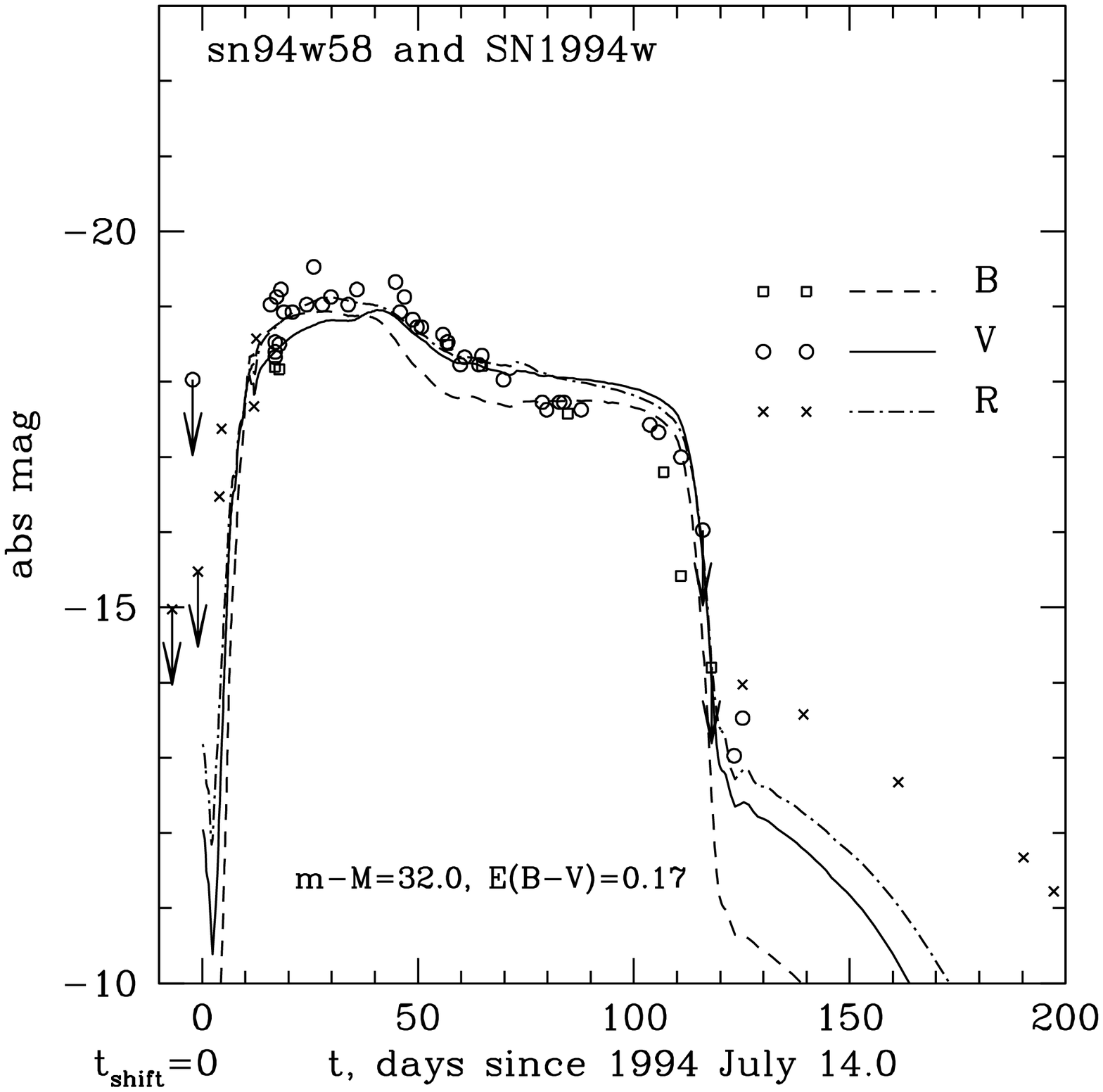} % {bvr94w58.ps}
  \includegraphics[width=0.45\textwidth]{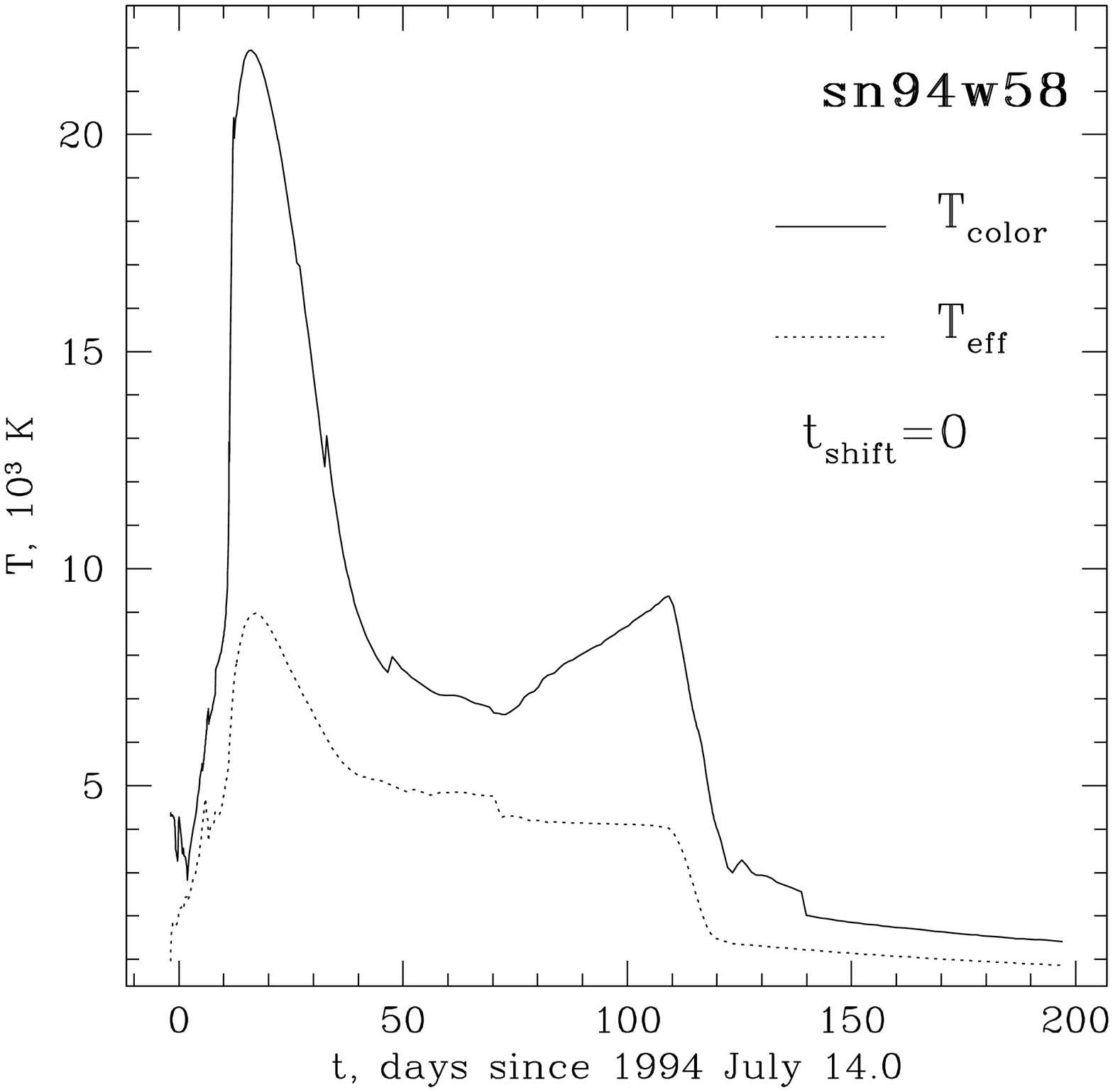}} % {Tph94w58.ps}}
  \caption{$BVR$ light curves and $T_\mathrm{ph}$ for the run sn94w58.}
  \label{blinn.sn94w58}
\end{figure}

Main features of SN 1994W
are well reproduced by the radiative shock propagating
at relatively low speed for several months in the
circumstellar shell. This shock forms its unique  lightcurve
with a plateau quite different from a classical SNeII-P.
(In the latter $UBV$ light curves diverge at the end of $V$ plateau,
while here all colors converge.)
The sudden drop of V-flux by $\sim 3.5$ magnitudes in 12 days
is also explained quite naturally.
The results presented in Fig.~\ref{blinn.sn94w58} are computed for the
supernova  model with ejected mass $7 M_\odot$ and huge radius of
presupernova $R_0=2\times10^4 R_\odot$ surrounded by a shell with
the density $\rho_\mathrm{wind}=12/r$ in CGS units.
%The wind [MAY BE NOT WIND] - Nikolai
%is cut at $R_\mathrm{cut}=2\times10^4 R\odot$.
%This dense and unsteady wind may be produced by relatively weak explosions
%prior to the collapse of the core \cite{blinn.Woosley86,blinn.GrNad91}.
The CS envelope has an outer cut-off radius $R=6.6\times10^4 R_\odot$.
The ejection of this dense extended CS envelope
%and unsteady wind
might be related to weak explosions occurring several years
prior to the collapse of the core \cite{blinn.Woosley86} (see also
\cite{blinn.GrNad91}). % -- here an explosion forming the "wind" just before collapse

\bigskip

%We conclude that photon spectra at shock breakouts in SNe~Ib/c
%peak near $3kT \sim 1$ keV and the X-ray fluences predicted can be
%observed by future missions up to tens of Mpc.

%\subsection*{Acknowledgements}

We are grateful to W.Hillebrandt and to the MPA staff for permanent and generous
cooperation. Support of MPA visitor program and of grants  NSF AST-97 31569,
NASA - NAG5-8128, RBRF 00-02-17230, RBRF 02-02-16500, the
Wenner-Gren Science Foundation, and the Royal Swedish Academy is acknowledged.

%INDEX%%%%%%%%%%%%%%%%%%%%%%%%%%%%%%%%%%%%%%%%%%%%%%%%%%%%%%%%%%%%%%%
% Please check with the editor of your book whether he plans to
% include a "mutual" subject index - if so, please code your entries
% in the standard syntax. For your own purposes you may print your
% "personal" index by using the following commands:
%
%\clearpage
%\addcontentsline{toc}{section}{Index}
%\flushbottom
%\printindex
%%%%%%%%%%%%%%%%%%%%%%%%%%%%%%%%%%%%%%%%%%%%%%%%%%%%%%%%%%%%%%%%%%%%%

\end{document}